\documentclass[pra,twocolumn,floatfix,showpacs,aps]{revtex4}
\usepackage[dvips]{graphicx}
\usepackage{amssymb,amsmath,bm}

\newenvironment{lcase}{\left\lbrace \begin{aligned}}{\end{aligned}\right.}

\begin{document}

\title{Orientation control by flow: Exact results and Langevin simulations}

\author{C. Tannous}
\affiliation{Laboratoire de Magn\'etisme de Bretagne CNRS-FRE 3117, \\
Universit\'e de Bretagne Occidentale, 29285 Brest, France.}

\date{April 2008}

\begin{abstract}
Rod shaped objects suspended in a flowing liquid might be orientated by the
velocity, nature of the liquid, the flow and the geometry of the channel containing the flow.
Orientation settings might  enhance or inhibit certain chemical reactions between the objects,
other chemicals or with the walls of vessels holding the flowing suspension.
The probability density function (PDF) describing the orientations of rod shaped objects 
in a flowing liquid  satisfies a Fokker-Planck equation whose solution is obtained analytically as 
well as numerically from Langevin simulations for different flow parameters. 
The analytical and numerical methods developed in the present work enable us to calculate
accurately the PDF for a range of the Peclet number $\alpha$ covering several orders of magnitude, 
$10^{-4} \le \alpha \le 10^{8}$. We apply these results to the experimental determination of dichroism
and birefringence of the suspension as a function of $\alpha$.


\end{abstract}


\maketitle

\section{Introduction}
In many physical, chemical, biological processes, the behavior and orientation of 
rod shaped objects  (RSO) such as fibers, nanotubes, nanowires, DNA, macromolecules...
suspended in a flowing liquid affect the transport, rheology, chemical and hydrodynamic 
characteristics of the suspension~\cite{Bird}. Orientation control for the sake of aligning or separating RSO, enhancing 
reaction between  them, with other chemicals or vessel walls is important in many areas of science and technology
such as sedimentation, blood flow, pulp and paper, polymer processing, microfluidic devices, ferrofluids...

Objects suspended in flow undergo two types of motion: smooth motion due to the average fluid velocity field and
erratic random motion produced by the fluctuating fluid velocity, temperature and inertia driven motion. 
The resulting change in the suspension 
microstructure can have a significant effect on the mechanical, thermal, optical, electrical,
magnetic and chemical properties.

The objective of this work is to investigate
the effect of non-turbulent flow on the rotational diffusion of a dilute suspension of 
non-interacting RSO in a planar contraction and how its impact might be measured with 
an optical measurement such as dichroism or birefringence of the suspension.

Generally a Fokker-Planck equation accurately models the orientation state of non-interacting
RSO in hydrodynamic nonhomogenous flow and the main orientation parameter is the rotational Peclet number $\alpha$
that represents the interplay between the randomizing effect of temperature induced rotational 
diffusion and the orienting effect of streamwise mean rate of strain due to the flow~\cite{Parsheh}.

 Two types of flow are of particular importance: simple shear flow and elongational 
(or, extensional) flow. Simple shear flow is a velocity profile where the gradient in the fluid
 flow velocity is constant, whereas for elongational flow the sample is compressed in one 
direction and elongated in the other direction. Such flows can be either stationary or oscillatory.

\begin{figure}[htbp]
\centering
\scalebox{1.2}{\includegraphics[angle=0]{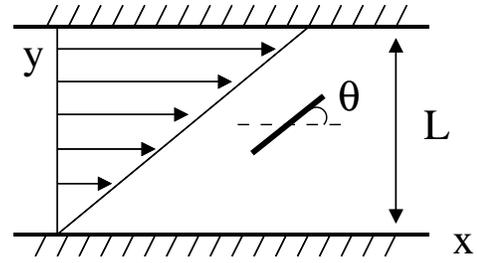}}
\caption{Geometry of the shear flow and the RSO of length $\ell$ making an angle $\theta$ with 
the flow direction along $x$. The shear rate of the flow is $\dot{\gamma}=\frac{\partial v_x}{\partial y}$.} \label{shear}
\end{figure}

The rotational Peclet number measuring the relative strength of hydrodynamic interactions and Brownian 
forces is defined mathematically as the ratio of hydrodynamic shear flow 
and rotational diffusion constant:

\begin{equation}
\alpha= \frac{\dot{\gamma}}{D_{rot}} 
\end{equation}

where  $\dot{\gamma}$ is the hydrodynamic shear rate and $D_{rot}$  the thermal diffusion~\cite{Reif}.

The flow shear rate defined by $\dot{\gamma}=\frac{\partial v_x}{\partial y}$ 
with $v_x$ the velocity field in the $x$ direction of the flow (see fig.~\ref{shear}) 
is related to translational degrees of freedom whereas $D_{rot}$, the Brownian diffusion coefficient, 
governs the rotational motion of the RSO  around its center of mass and hence relates to rotational
degrees of freedom.

The rotational diffusion coefficient, $D_{rot}$ is given by~\cite{Doi}:

\begin{equation}
D_{rot}=\frac{3 k_B T }{ \pi \eta_s \ell^3} [\ln (\frac{\ell}{d})-0.8]
\end{equation}

with $k_B$ the Boltzmann constant, $T$ the temperature, expresses the influence of flow
on orientation. Usually it is determined experimentally from RSO orientation studies in a 
constriction with flat walls. $\ell$ is the RSO length  $d$ its diameter
and $\eta_s$  the solvent viscosity.

Non-interacting dilute concentration of RSO of density $\rho$, yield the conditions:
$\rho \ell^2 d < 1$ and $ \rho \ell^3 < 1$.

The Reynolds number defined as~\cite{Landau}: 
$Re=\frac{\rho_s \dot{\gamma} L^2}{\eta_s}$ is much smaller than 2000 in order to have a non-turbulent
flow. $\rho_s$ is solvent density and $L$ the width of the planar contraction (see fig.~\ref{shear}).

Boeder~\cite{Boeder} is the first to have studied this problem in the bulk of a flowing liquid from the
 theoretical point of view. He derived, without the use of the Langevin formalism or the 
Fokker-Planck equation~\cite{Gardiner}, an ordinary differential equation (ODE) governing the 
PDF $P(\theta)$ describing the average orientations of the RSO:

\begin{equation}
\frac{d^2}{d\theta^2} P(\theta) + \frac{d}{d\theta} [\alpha \sin^{2}(\theta) P(\theta)] =0
\label{boeder}
\end{equation}

The above ODE is derived for the motion of RSO, of negligible 
cross-sectional area, in the plane of the flow, without any boundary conditions and in the dilute case. 
The angle $\theta$ describes the orientations of the RSO with respect to the liquid flow
direction (see fig.~\ref{shear}. 

Since the PDF $P(\theta)$ is $\pi$-periodic one might write a  Fourier series solution valid for small and 
large values of $\alpha$. For small values of 
$\alpha$ we perform a perturbation analysis of the Fourier series whereas for large 
values of $\alpha$ asymptotic analysis might be done~\cite{Boeder}.
Many improvements have been made since,
to remove restrictions on the cross-sectional areas of the RSO, 
consider rotational diffusion in three dimensions  and introduce internal vibrational and rotational  degrees of freedom
within the RSO. 

The purpose of this work is to provide a solution in closed form for the above ODE with different analytical and 
numerical methods to obtain $P(\theta)$ for a wide range of 
the Peclet number $\alpha$  and gauge the impact on orientation control and the measurable optical properties.
Another goal is to compare the analytical approach with Langevin simulations of the PDF, under the
same conditions in the bulk of a flowing liquid. This comparison is useful because it can confirm
on one hand the robustness of the analytical methods. On the other hand it provides a necessary limiting bulk condition for 
similar simulations near solid surfaces.  

This paper is organised as follows: In Section II, we present an exact analysis of the
ODE (see eq.~\ref{boeder}) to obtain the probability distribution function (PDF), for a wide range of $\alpha$. 
In Section III, Langevin simulation and comparison with the exact results are presented.
In section IV, large values of $\alpha$ are treated along with scaling results, 
optical properties and flow control. The conclusions are given in Section V.

\section{Analytical solution}
 In this section, a procedure for the accurate numerical analysis of the
ODE (see eq.~\ref{boeder}) and its associated probability distribution function, $P(\theta)$, 
is given for a wide range of $\alpha$. Turbulence effects are known to take place for large values of $\alpha$ 
(typically for $\alpha \ge 10^{4}$).
 Although the above ODE (eq.~\ref{boeder}) ceases to apply  beyond this limit, 
the present mathematical analysis is not limited by this, and numerical solutions may be calculated in 
our approach for values of $\alpha \sim 10^{8}$. $P(\theta)$ is the solution of a second-order ODE 
(eq.~\ref{boeder}). The PDF is $\pi$-periodic since the RSO
 are indistinguishable when oriented at $\theta \mbox{ or } \theta+\pi$; hence the $\pi$-periodic boundary conditions:

\begin{equation}
P(0)=P(\pi)   \mbox{  and   }  P'(0)= P'(\pi)  
\label{BVP}
\end{equation}

where  $P'(\theta)=\frac{d}{d\theta}P(\theta)$.
Moreover, the PDF has to be normalised over the interval $[0,\pi]$:

\begin{equation}
\int^{\pi}_{0}{P(\theta)d\theta}=1
\label{norm}
\end{equation}

The determination of the associated PDF is consequently a constrained (because of the normalization condition)
boundary value problem (BVP) eq.~\ref{boeder}. 
Nevertheless one may derive two other versions of the ODE eq.~\ref{boeder}. 
The first is a 1D equation that takes the form:

\begin{equation}
P'(\theta)+ \alpha \sin^{2}(\theta)P(\theta) =C
\label{IVP1}
\end{equation}

with initial condition: $P(\theta=0)=P(0)$.
This is mathematically sound provided the initial value $P(0)$  and the constant $C$ are known as shown below. 
The second form is a 2D system that may be presented as:

\begin{equation}
\begin{lcase}
\frac{d}{d\theta}P(\theta) & = & P'(\theta) \hspace{6cm} \\
\frac{d}{d\theta}P'(\theta) & = & -\alpha \sin(\theta)[\sin(\theta)P'(\theta)+2\cos(\theta)P(\theta)]  \hspace{1cm}
\end{lcase}
\label{IVP2}
\end{equation}

Several technical procedures may be used to determine $P(\theta)$, depending on the different possible
 formulations of the problem, as an initial value problem (eq.~\ref{IVP1} and eq.~\ref{IVP2}) or as a BVP (eq.~\ref{BVP}).
 In all cases two quantities, the constant $C$ in eq.~\ref{IVP1} (equal to $P'(0)$ by substituting 
$\theta=0$ and assuming finiteness of $P(\theta \rightarrow 0 )$ in eq.~\ref{IVP1}) and the initial value $P(0)$, 
have to be evaluated for any  value of $\alpha$. We have developed two methods, described
 below, to determine $C$ and $P(0)$. Firstly, a direct method based on the solution of eq.~\ref{IVP1}, and 
secondly a minimisation method based on a multidimensional secant method (also called Broyden method~\cite{Recipes}).
 
The direct  method to evaluate $C$ and $P(0)$  is as follows.
 The formal solution of eq.~\ref{IVP1} may be given generally as:

\begin{eqnarray}
P(\theta) = C \exp[\alpha [\sin(2\theta)/2-\theta]/2] \times \hspace{3cm} \\ \nonumber
\hspace{2cm}  \int_{-\infty}^{\theta} \exp[-\alpha(\sin(2x)/2-x)/2]dx
\label{unstable}
\end{eqnarray}

The lower limit  $-\infty$ is surprising since the problem is defined over the angular interval $[0,\pi]$.
It can analytically be shown that the 
lower limit  $-\infty$ is the only possibility compatible with the boundary conditions given by eq.~\ref{BVP}. 
Performing a change of variables, the solution may then be written as:

\begin{eqnarray}
P(\theta) = (2C/\alpha) \hspace{1mm} \exp[\alpha \hspace{1mm} \sin(2\theta)/4]  \times  \hspace{3cm} \\ \nonumber
\int_{0}^{\infty} \exp(-x) \exp[ \hspace{1mm}\alpha \hspace{1mm} \sin(4x/\alpha-2\theta)/4]dx
\label{stable}
\end{eqnarray}
 
The form in eq.~\ref{stable} is more stable numerically than the previous one of eq.~\ref{unstable}, since the 
numerically troublesome $\exp(\theta)$   and $\exp(-\theta)$ terms are avoided. Nevertheless when $\alpha$ 
increases the $\exp[\alpha \hspace{1mm} \sin(2 \theta)/4]$  term will cause problems despite the bounded values 
of the sine function, forcing us to turn to other methods based on extrapolation techniques. 
The constant $C$ is next determined from the normalisation condition of the PDF (eq.~\ref{norm}) yielding:

\begin{eqnarray}
C = \hspace{8cm} \nonumber  \\
\frac{(\frac{\alpha}{2}) }{ \int_{-\frac{\pi}{2}}^{\frac{\pi}{2}} d\theta \int_{0}^{\infty} dx \hspace{1mm} \exp(-x)  
\hspace{1mm} \exp[(\frac{\alpha}{2}) \sin(\frac{2x}{\alpha}) \cos(2\theta -\frac{2x}{\alpha} )]} \nonumber \\
 \hspace{8cm}
\label{eq:C}
\end{eqnarray}

whereas $P(0)$   is given by

\begin{equation}
P(0) = (2C/\alpha) \int_{0}^{\infty} \exp(-x) \hspace{1mm} \exp[\hspace{1mm} \alpha \hspace{1mm} \sin(4x/\alpha)/4]dx
\label{eq:P}
\end{equation}

The PDF depends on $\alpha$ which we may want to vary over several orders of magnitude. 
The difficulty in solving the ODE stems from the fact that its nature may be modified 
when $ \alpha$ increases, turning the problem into a singular perturbation one in  $ \alpha^{-1}$ as
detailed in section IV.

\begin{figure}[htbp]
\centering
\scalebox{0.3}{\includegraphics[angle=-90]{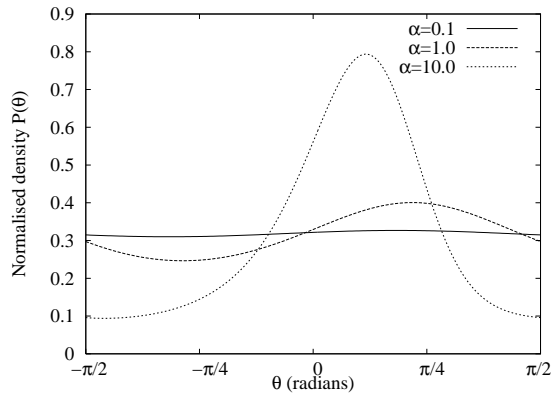}}
\caption{Analytical PDF as a function of $\theta$ for small $\alpha$=0.1, 1. and 10.0 normalised
over the interval $[-\pi/2,\pi/2]$} \label{small1}
\end{figure}

\section{Langevin simulations and comparison with the exact equation}
We convert the deterministic equation of motion for the RSO 
into a Langevin equation by adding a random force term and analyze statistically
the resulting equation.

We start from the angular speed:

\begin{equation}
\frac{d\theta}{dt}=\omega= -\dot{\gamma}\sin^{2}(\theta)
\label{angspeed}
\end{equation}

and add to it a random term $\lambda \xi(t)$ with White noise statistical properties: 

\begin{equation}
<\xi(t)>=0 \mbox{  and  } <\xi(t)\xi(t')> = \delta(t-t') 
\end{equation}

This means the hydrodynamic forces tend to act on 
the RSO rotating them in the shear flow with an average angular speed $\omega$
in the presence of perturbating fast random orientations represented by $\xi(t)$.
The equation of motion for $\theta$ becomes:

\begin{equation}
d\theta=  -\dot{\gamma}\sin^{2}(\theta) dt + \lambda dW(t)
\end{equation}

where $dW(t)=\xi(t)dt$ is a Wiener process~\cite{Gardiner}.

This is an  It\^o Stochastic Differential Equation (SDE) (see for instance Gardiner~\cite{Gardiner})
that can be transformed into a 1D Fokker-Planck equation that will be identified with
the ODE equation~\ref{boeder}.

A single stochastic variable $\theta$ It\^o SDE~\cite{Gardiner} of the form:

\begin{equation}
d\theta= A(\theta,t) dt + \sqrt{B(\theta,t)} dW(t)
\end{equation}

is equivalent to a 1D Fokker-Planck equation:

\begin{equation}
-\frac{d }{d\theta}[A(\theta) P(\theta)] + \frac{1}{2} \frac{d^2 }{d\theta^2}[B(\theta) P(\theta)] =0 
\end{equation}

\begin{figure}[htbp]
\centering
\scalebox{0.3}{\includegraphics[angle=-90]{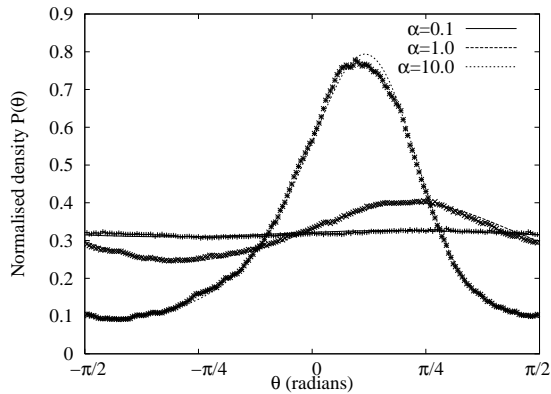}}
\caption{Langevin simulated PDF as a function of $\theta$ for small $\alpha$=0.1, 1. and 10.0 normalised
over the interval $[-\pi/2,\pi/2]$. The analytical curves are displayed for comparison.} \label{small2}
\end{figure}

Identification with the ODE equation~\ref{boeder} yields the value of the coefficient $A(\theta)$ as:
$A(\theta)=-\frac{1}{2}B \alpha \sin^{2}(\theta)$ whereas $B$ is a constant independant of $\theta$.
Hence the SDE writes:

\begin{equation}
d\theta= -\frac{1}{2}B \alpha \sin^{2}(\theta) dt + \sqrt{B} dW(t)
\end{equation}

The hydrodynamic forces tend to act on the RSO rotating them in the
shear flow with an average angular speed $\omega$, given by eq.~ref{angspeed}.

This completely characterizes the SDE parameters as:

\begin{equation}
\dot{\gamma}= \frac{1}{2}B \alpha, \hspace{2cm} B= 2 D_{rot}
\end{equation}

The method of integration we use, is a 4th order
Stochastic Runge-Kutta routine: 3O-3S-2G developed by Helfand~\cite{Helfand}
where 3O is third order, 3S is three stages 2G means we need 
two Gaussian Random Variables per step.
Order $k$ means the solution agrees with the average Taylor expansion~\cite{Greiner}
of the series solution to order $h^k$ where $h$ is the integration step.

In order to assess the validity of the method, we ran several tests and compared
the results to known analytical solutions.

The results we find for the Langevin simulated PDF are displayed
in fig.~\ref{small2} and fig.~\ref{large2} for small and large values of the
Peclet number. The results of the Langevin simulation converged after $\sim 10^7$ steps
toward the analytically determined PDF.

\section{Large values of the Peclet number: scaling and optical properties}
 In order to cope with the wide range over which  $\alpha$ may
 vary, we classify the various methods for solving the problem according to the value of $ \alpha$.\\

For moderate $\alpha$ in the range : $[10^{-4}, 10 ^{2} ]$, we can proceed either directly from 
the analytic solution of the first-order ODE or once $ C$ and $P(0)$   are known,
 a simple 1D Runge-Kutta method is used to solve the first-order ODE. \\

In order to find $C$ and $P(0)$ by a minimisation
 method based on a multidimensional secant (Broyden method), and a 2D Runge-Kutta integration
 method to solve the two first-order ODE system (Eqs.~\ref{IVP2}). 
We have the following two conditions:

\begin{eqnarray}
\mbox{ minimum  }   |P(0)-P(\pi)|;   \nonumber \\
\mbox{ minimum  }   |\int_{0}^{\pi}P(\theta) d\theta -1|
\label{min}
\end{eqnarray}

In the above minimisation problem, the first condition comes from periodicity of the PDF, whereas the second  
expresses the normalisation of the PDF. \\

When  $\alpha$ becomes large i.e. in the range : $[10^{2}, 10 ^{5} ]$, we 
 calculate $ C$ and $P(0)$   by an extrapolation method and solve eq.~\ref{IVP1}, or the system~\ref{IVP2}, with singular 
perturbation integration methods~\cite{Recipes}. To illustrate the different numerical methods used in our approach to solve
 the Boeder differential equation, and to calculate the PDF, some numerical results are presented, for a
 wide range of values of $\alpha$. In Figs. 2 and 4,  these PDF functions are depicted, for moderate ($\alpha$=0.1,1,10.) 
and large values ($\alpha$=100., 1000. and 10,000) of $\alpha$. The PDF results are normalised with respect
 to unity. The scaling results are presented next for large values of $\alpha$.

\subsection{Scaling results} 

Eq.~\ref{eq:C} can be simplified for large $\alpha$ by 
replacing the cosine term in the exponential integrand by  1, since it will lead to
higher order terms in $1/\alpha$ and reducing the double integration appearing 
in the $C$ denominator into a simpler one, namely:

\begin{equation}
C = \frac{\alpha^{\frac{1}{3}} }{\left[  2 \theta_{max} \int_{0}^{\infty} \exp( -2 x^{3}/3 )dx \right]}
\end{equation}

where $\theta_{max}$ is the value of the angle that corresponds to the PDF maximum at which the
ODE eq.~\ref{boeder} writes:

\begin{equation}
\alpha \hspace{1mm} \sin^{2}(\theta_{max})P(\theta_{max})=C
\label{max}
\end{equation}

Using the definite integral~\cite{Gradstein}:

\begin{equation}
 \int_{0}^{\infty} x^{\nu-1} \exp( -\mu x^{p})dx=\frac{1}{p}\mu^{-\frac{\nu}{p}} \Gamma(\frac{\nu}{p})
\end{equation}

we get:

\begin{equation}
C = \frac{3 \alpha^{\frac{1}{3}} }{\left[  2 \theta_{max} {(\frac{2}{3})}^{-\frac{1}{3}} \Gamma(\frac{1}{3}) \right]}
\end{equation}

Similarly the argument of the exponential integrand in eq.~\ref{eq:P}, yields

\begin{eqnarray}
P(0) & = & (2C/\alpha^{\frac{1}{3}})  \int_{0}^{\infty} \exp( -8x^{3}/3\alpha^{2} )dx \nonumber \\
  & = & (2C/3 \alpha^{\frac{1}{3}}) {\left(\frac{8}{3}\right)}^{-\frac{1}{3}} \Gamma(\frac{1}{3})
\end{eqnarray}

\begin{figure}[htbp]
\centering
\scalebox{0.3}{\includegraphics[angle=-90]{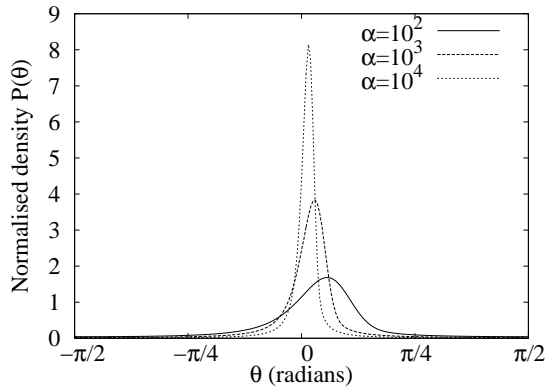}}
\caption{Analytical PDF as a function of $\theta$ for large $\alpha$=100.0, 1000. and 10,000. normalised
over the interval $[-\pi/2,\pi/2]$} \label{large1}
\end{figure}

The normalisation of the PDF in Eq.~\ref{norm} when $\alpha$ is large, yields after using eq.~\ref{max}:

\begin{equation}
 \theta_{max}P(\theta_{max}) \sim \mbox{constant}
\label{area}
\end{equation}

which is the area under the PDF curve since the distribution function becoming sharply 
peaked around $\theta_{max}$ , leads to a PDF approximately 
triangular in shape with a height $P(\theta_{max})$   and  a base equal to $2\theta_{max}$.
Hence, we obtain the following leading behaviours:

\begin{eqnarray}
P(0)  \sim  \alpha^{1/3},  \hspace{2mm} C \sim \alpha^{2/3}; \nonumber \\
\theta_{max}  \sim  \alpha^{-1/3},  \hspace{2mm} P(\theta_{max}) \sim \alpha^{1/3} 
 \end{eqnarray}

\begin{figure}[htbp]
\centering
\scalebox{0.3}{\includegraphics[angle=-90]{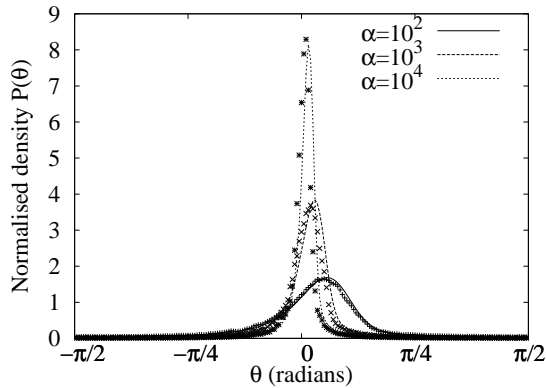}}
\caption{Langevin simulated PDF as a function of $\theta$ for large $\alpha$= 100.0, 1000. and 10,000. normalised
over the interval $[-\pi/2,\pi/2]$. The analytical curves are displayed for comparison.} \label{large2}
\end{figure}

The exponents controlling the asymptotic behaviour above are all consistent with respect
to each other, besides we have checked them numerically up to quite large values of  $\alpha \le 10^{8}$ . 

Recently, Fry et al.~\cite{Fry} made optical studies of Carbon nanotube suspensions in polyisobutylene
and aqueous single-stranded DNA solutions and were able to show that shear-induced birefringence and dichroism
(SABD) $\Delta n', \Delta n"$
are both proportional to a single quantity $S=<P_2(\cos \theta)>$ so-called nematic order parameter
that is zero for an isotropic distribution of orientations and 1 for perfect alignment. $P_2(\cos \theta)$ is
the Legendre polynomial of order 2 and $\Delta n'$ is extracted from the phase difference between
the transmitted and reference signal whereas $\Delta n"$ is extracted from the attenuation of
light intensity.

They found that SABD scale as $\alpha^{0.16}$ for values of 
the Peclet number as large as $10^{10}$ as they approach the saturation value of 1.

\begin{figure}[htbp]
\centering
\scalebox{0.3}{\includegraphics[angle=-90]{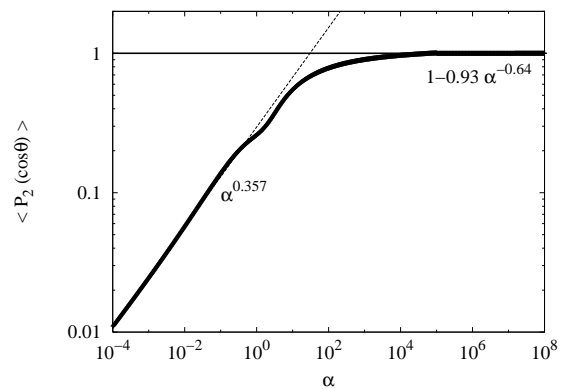}}
\caption{Average value of $P_2(\cos \theta)$ versus Peclet number $\alpha$ 
displaying the scaling behaviour at large values of $\alpha$ as it is approaching 1.
The power law  behaviour $\sim \alpha^{0.357}$ for small values of $\alpha$ (straight dotted lines) and 
$\sim 1- 0.93 \alpha^{-0.64}$ for large values of $\alpha$ are indicated.} \label{dichro}
\end{figure}

Therefore we set out to analyse the scaling of $S$ in the Boeder case. Our results shown in fig.~\ref{dichro}
indicate that in our case, the SABD saturate to 1 and that might be attributed to the
following argument.

As the Peclet number increases the orientational PDF becomes sharply peaked around $\theta_{max}$
that scales as $\alpha^{\frac{1}{3}}$. Since, it is required that the PDF should be normalised 
we ought to have the value of $S$ behaving as:

\begin{eqnarray}
S &= & \int_{-\frac{\pi}{2}}^{\frac{\pi}{2}} P(\theta)    P_2(\cos \theta) d\theta  \nonumber  \\
 & \sim  &  \int_{-\theta_{max}}^{\theta_{max}}  P(\theta)   P_2(\cos \theta) d\theta \nonumber  \\
   & \sim  &  2 \theta_{max}  P(\theta_{max}) P_2(\cos \theta_{max})
\end{eqnarray}

That implies, given eq.~\ref{area}, that $S$ behaves as $P_2(\cos \theta_{max})= \frac{1}{2} (3 \cos^2 \theta_{max}  -1)
 \sim 1 -\frac{3}{2} \alpha^{-\frac{2}{3}}$ close to the behaviour found in fig.~\ref{dichro} and obtained from full 
numerical integration.

Since  $\theta_{max}$ is a measure of the standard deviation (for large values of $\alpha$) in the
fluctuations of the angle $\theta$ about the flow main direction, the RSO orientation controllability
with flow can be analysed with the variation of the angle  $\theta_{max}$ with the Peclet number $\alpha$.
Increasing $\alpha$ makes the orientational PDF more sharply peaked around $\theta_{max}$ and 
with the requirement that the PDF should be normalised, $\theta_{max}$ approaches the 
standard deviation. 

\begin{figure}[htbp]
\centering
\scalebox{0.3}{\includegraphics[angle=-90]{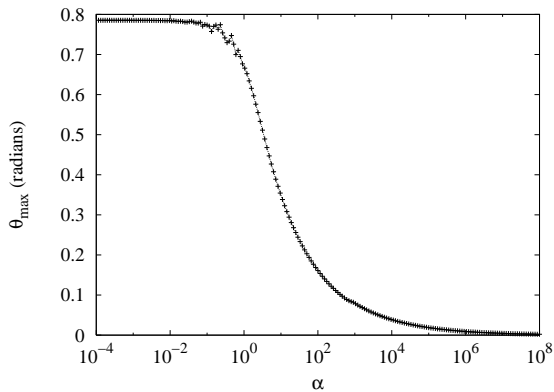}}
\caption{Variation of the angle $\theta_{max}$ (in radians) with the Peclet number $\alpha$. For small values of $\alpha$,  
 $\theta_{max}$ saturates at a value of about 0.8 radian that is close to $\frac{\pi}{4}$ as seen in fig.~\ref{small1}. When
$\alpha$ is increased beyond several 100, $\theta_{max}$ drops to small values indicating that orientation control has set in. }
 \label{control}
\end{figure}

In fig.~\ref{control} the variation of  $\theta_{max}$ with the Peclet number $\alpha$
is depicted implying that orientation controllability is achieved when $\theta_{max}$
drops below a small specified value. For instance, one has to have $\alpha \sim 400$ in order to
achieve $\theta_{max} \le 0.1 $ radian (see fig.~\ref{control}) whereas $\alpha \sim 6\times 10^{5}$ is required 
in order to have $\theta_{max} \le 0.01 $ radian.

\section{Conclusions}
 The PDF describing the average orientations of RSO in a flowing liquid is analytically  
determined and accurately evaluated as a function of $\alpha$, the rotational Peclet number 
over a wide range. The impact of $\alpha$ on orientation control and optical properties is examined.
Special analytical as well as numerical methods are developed and presented in order to calculate accurately 
this PDF for a wide range of $\alpha$ over the interval $ [10^{-4} - 10^{8}]$.
Langevin simulations agree well with the 
numerical solutions of the PDF in the bulk of the flowing liquid for arbitrary values of $\alpha$
and confirm the validity of the analytical solution.
 Scaling results (valid for $ \alpha \ge 10^{3}$) are also  presented. The mathematical
 nature of the ordinary differential equation that the PDF should satisfy is revealed as 
a singular perturbation problem  when $\alpha^{-1}$ becomes smaller than about $10^{-3}$ and special
numerical techniques~\cite{Recipes} ought to be used in order to evaluate the PDF and related physical quantities.\\
The analytical approach, Langevin simulation and scaling results should be considered equally as complementary 
tools for this type of study and as a valuable guide to the harder cases where we have either turbulent flow,
presence of RSO internal degrees of freedom or more complex flow geometries.

{\bf Acknowledgements}\\
The author would like to thank M. Aoun for his help in the statistical analysis
 of the simulation results.

\end{document}